\documentclass[preprint,12pt]{elsarticle}
\usepackage{graphics}
\usepackage{amssymb}
\usepackage{lineno}
\journal{Nuclear Physics A}
\begin{document}
\begin{frontmatter}
 
\author[q]{Tina Pollmann\corref{cor1}} 
\ead{tina@owl.phy.queensu.ca} 

\author[q]{Mark Boulay}
\author[q]{Marcin Ku\'zniak}

\cortext[cor1]{Corresponding author} 
\address[q]{Department of Physics, Engineering Physics, and Astronomy \\
Queen’s University, Kingston, Ontario, K7L 3N6, Canada}
\title{Scintillation of thin tetraphenyl butadiene films under alpha particle excitation}
\author{}
\address{}

\begin{abstract}
The alpha induced scintillation of the wavelength shifter 1,1,4,4-tetraphenyl-1,3-butadiene (TPB) was studied to improve the understanding of possible surface alpha backgrounds in the DEAP dark matter search experiment. We found that vacuum deposited thin TPB films emit 882~$\pm$210 photons per MeV under alpha particle excitation. The scintillation pulse shape consists of a double exponential decay with lifetimes of 11$~\pm$5~ns and 275~$\pm$10~ns.
 
\end{abstract}

\end{frontmatter}
    \linenumbers

\section{Introduction}
A number of noble liquid based dark matter direct detection experiments, such as DEAP\cite{deap}, MiniClean\cite{McKinsey:2007p10091} and ArDM\cite{ArDM}, rely on thin films of 1,1,4,4-tetraphenyl-1,3-butadiene (TPB) coated on the detector walls to shift the wavelength of the noble gas scintillation light from the vacuum ultra-violet to a peak wavelength of 420 nm\cite{Burton:1973p6267}, which can then be detected by conventional photomultiplier tubes (PMTs). However, TPB also scintillates, which may lead to background signals from omnipresent alpha emitters embedded in the detector wall material or the TPB layer itself, compromising the sensitivity of the experiment.

We present a study of the alpha induced scintillation of thin films of pure scintillation quality TPB, emphasizing the light yield and the pulse shape of the scintillation signals. This work was carried out as an R\&D project for the DEAP\cite{deap} experiment.

\section{TPB evaporation system}
\subsection{Apparatus}
\begin{figure}[htbp] 
   \centering
   \includegraphics[]{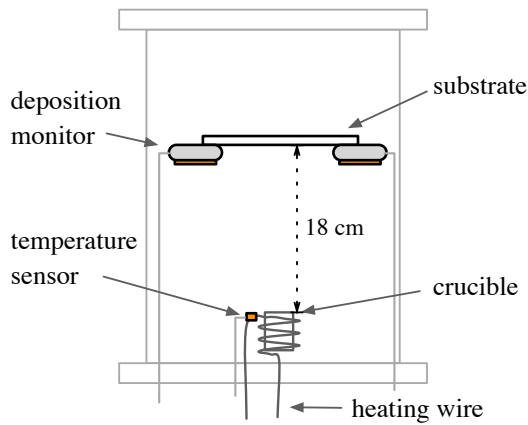} 
   \caption{Diagram of the TPB evaporation system.}
   \label{fig:evap}
\end{figure}

TPB films of thicknesses in the order of 2-4~$\mu$m were produced by evaporating crystalline scintillation quality TPB powder from American Chemicals LTD under vacuum onto glass or polished acrylic plates. The TPB was evaporated from a quartz glass crucible 1~cm high and 1~cm in diameter located at the bottom of the chamber. The crucible was heated to the target temperature of approximately 200$^{\circ}$~C by means of a heating wire wrapped around it (see figure~\ref{fig:evap}). An OMEGA resistance temperature sensor was clipped to the heating wire near the crucible and its readout was coupled to the power supply of the heater to maintain a constant temperature. The glass or acrylic plates were held on a rack 17.8$\pm$0.3~cm above and centred on the crucible. Installed at the same height and on each side of the rack were two quartz crystal deposition monitors to monitor the thickness of the deposited film.

\subsection{Method of evaporation} 
The glass plates (radius 6~cm) and acrylic plates (radius 10~cm, thickness 1.9~cm) were prepared for evaporation by the following steps: 
\begin{enumerate}
\item Washed with tap water and detergent and rinsed with tap water.
\item Wiped with isopropanol.
\item Wiped with a mixture of 20\% ethanol and 80\% ultra pure water.
\item Wiped with aluminum foil. \label{list:tin}
\item Rinsed with ultra pure water.
\item Blown dry with nitrogen or argon gas.
\end{enumerate}
Step \ref{list:tin} was introduced after observing that the TPB distribution on the cleaned plates showed signs of the surface treatment, possibly due to surface charges accumulated during wiping. The TPB surface on all substrates used for these measurements was very smooth and powdery white under visual inspection. This indicated that the TPB formed either very small polycrystalline structures or did not crystallize at all. Coatings produced two months later with the same procedure again showed clear signs of surface treatment in the form of visible local thickness variations for yet unknown reasons.

To calibrate the deposition monitors, two test depositions on glass plates were performed evaporating 0.364$\pm$0.005~g and 0.255$\pm$0.005~g of TPB respectively. During those test depositions, small glass pieces were also installed at the same height of the plates to get TPB films at positions further out than the radius of the plates. The deposition on both glass plates was started when the vacuum in the evaporator was at 5~$\cdot 10^{-5}$~mbar. The pressure rose to up to 1~$\cdot 10^{-4}$~mbar during the evaporation process. 

For the measurement of the light yield, two TPB coatings were done. For coating A, 0.410$\pm$0.005~g of TPB was evaporated on an acrylic plate at 5~$\cdot 10^{-5}$~mbar pressure in the evaporator. An evaporation temperature of 212$^{\circ}$C was necessary to evaporate all the TPB in the very packed crucible. For coating B, 0.284~g TPB was evaporated at a pressure of 2~$\cdot 10^{-4}$~mbar and a temperature of 200$^{\circ}$C.

\section{Deposition monitor calibration}
\begin{figure}[htbp] 
   \centering
   \includegraphics[width=0.8\textwidth]{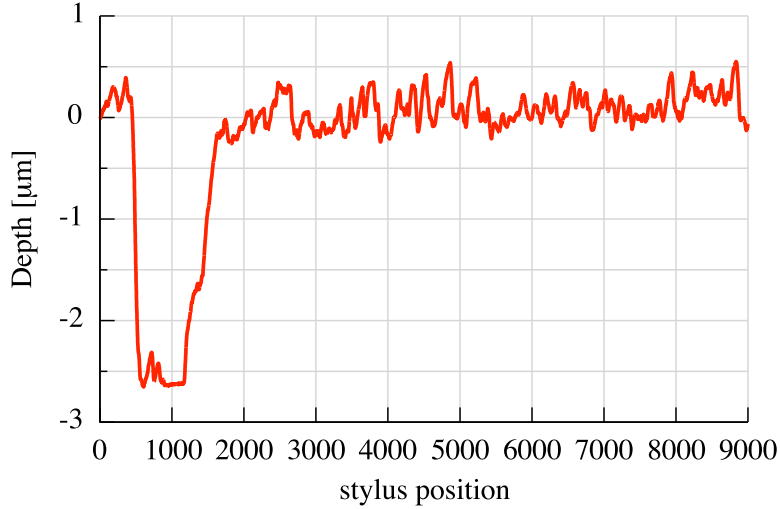} 
   \caption{Typical profile meter scan of a scratch in the TPB film. The TPB thickness was determined by subtracting the average depth of the valley (stylus position 800 to 1200) from the average height of the TPB (position 1500 to 9000). The length of the scan is 500~$\mu$m  }
   \label{fig:scratch}
\end{figure}

The thickness of the TPB film across the substrate is a function of the radial distance from the axis of the crucible. To correlate the deposition monitor readings with the thickness at the position of the light yield measurement, the TPB film on the glass substrate was scratched with a scalpel and the depth was measured using a Dektak 8M stylus profile meter from Veeco Instruments. A sample scratch profile is shown in figure \ref{fig:scratch}. There was no indication that the glass beneath the TPB was damaged. 

\begin{figure}[htbp] 
   \centering
	\includegraphics[width=0.8\textwidth]{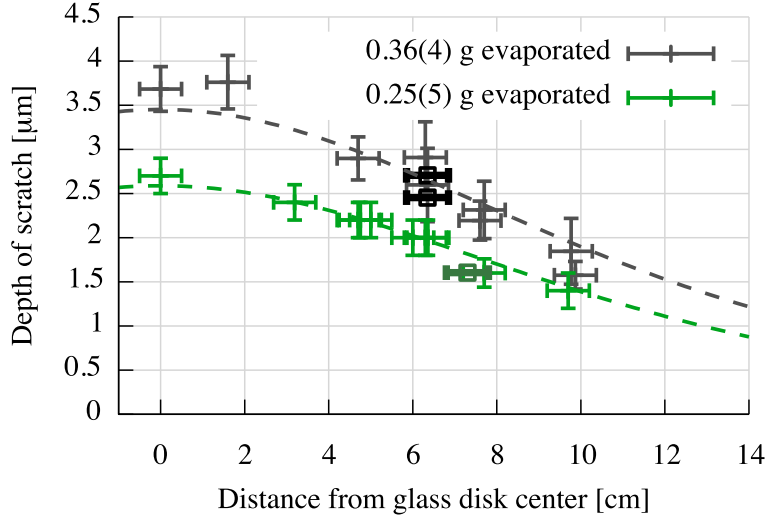}
   \caption{The variation in TPB thickness with radial distance from the crucible axis for the two test depositions on glass. The light thin points are results from the profile meter measurement, while the darker bold boxes are the readings from the deposition monitors. Dashed curves are best fits with the distribution from ref.~\cite{Chaleix:1996p10018}.}
   \label{fig:depth}
\end{figure}

The scratch depths across the substrates, shown in figure \ref{fig:depth}, show a radial thickness variation consistent with the distribution developed by Chaleix et al. \cite{Chaleix:1996p10018}: 
\begin{equation}
\nu_{d} = \frac{\nu_{e}}{\rho} g(\theta) \frac{d}{(r^{2} + d^{2})^{3/2}}  \label{eq:distr}
\end{equation}
where $\nu_{e}$ and $\nu_{d}$ are the rates of evaporation and deposition, d is the source-substrate distance, r is the radial distance from the axis of the crucible to a point on the substrate and $\rho$ is the density of the TPB film. The geometric factor $g(\theta)$ depends on the evaporation conditions. For conditions with a volume right above the crucible where the probability of interaction between TPB molecules is very high, followed by a volume where the molecules move in straight lines, the geometric factor is given by\cite{Chaleix:1996p10018} 
\begin{equation}
g(\theta) = \frac{1}{\pi} \frac{e^{\beta^{2}} (1 + \mbox{erf}(\beta)) ( \frac{3\beta \sqrt{\pi}}{2} + \frac{\beta^{3}}{\pi}) + \beta^{2} + 1 }{1 + \frac{u}{\alpha}\sqrt{\pi}e^{ \frac{u^{2}}{\alpha^{2} }} (1 + \mbox{erf}( \frac{u}{\alpha})) } \mbox{cos}(\theta)
\end{equation}
where
\begin{equation}
\beta = \frac{u}{\alpha} \mbox{cos}(\theta)
\end{equation}
\begin{figure}[htbp] 
   \centering
   \includegraphics[width=0.35\textwidth]{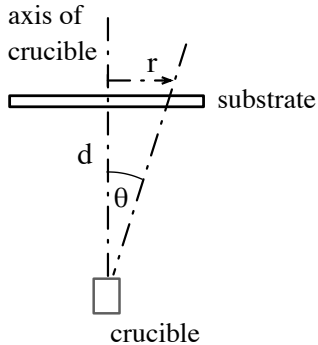} 
   \caption{Geometry of the TPB evaporation system.}
   \label{fig:geometry}
\end{figure}
and $\theta$ is the angle between the axis of the crucible and a line from the crucible to a point on the substrate, as shown in figure \ref{fig:geometry}.
The fit paramater $\frac{u}{\alpha}$, which depends on the vapour temperature at the boundary where the TPB molecules start moving without collisions, was found to be 0.284$\pm$0.013 for the substrate with 0.364~g TPB evaporated and 0.224$\pm$0.004 for the substrate with 0.255~g TPB evaporated. 

The density of the TPB film enters into the mass distribution equation (eq.~\ref{eq:distr}). It was fitted from the scratch measurement data to be 1.16$\pm$0.04~g/cm$^{3}$. The uncertainty takes into account the additional constraint that the deposition monitor readings, which also depend on the density of the TPB layer, had to be within errors of the scratch test measurement. This result is also consistent with the TPB density estimated from the mass of the substrate before and after evaporation.

Since the deposition on sample A was done at a different temperature, the parameter $\frac{u}{\alpha}$ differed significantly from the above. It was fitted from the deposition monitor readings to be 0.36$\pm$0.02 and the function was then used to calculate the TPB thickness at all locations where scintillation data was taken. The readings on the deposition monitors for the coating of sample B were not consistent with the above thickness distribution function, but instead indicated a uniform thickness distribution, as would be expected for an evaporation pressure high enough to not allow for a region of collision-less movement of the TPB molecules.

\section{TPB scintillation measurement}  

\begin{figure}[htbp] 
   \centering
	\includegraphics[]{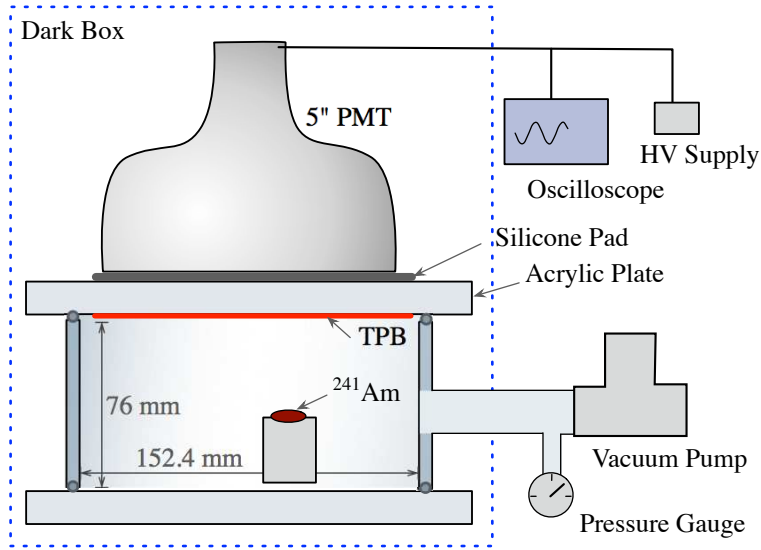}
   \caption{Diagram of the setup used to determine the light yield of a TPB film under alpha particle excitation.}
   \label{fig:chamber}
\end{figure}

Acrylic plates were used as a substrate for the light yield measurement because acrylic is also the material covered with TPB in the DEAP experiment.
The acrylic plates were installed in the scintillation chamber shown in figure~\ref{fig:chamber}. The chamber consists of a stainless steel cylinder sealed on the bottom with a plain acrylic plate and on the top with the TPB coated acrylic plate. It was viewed by a 5" PMT (ETL 9390B), optically coupled to the top of the acrylic plate with a 1~mm thick silicone pad.

A $^{241}$Am alpha source (5.49 MeV alpha energy) with a collimator of 44$^{\circ}$ opening angle could be placed at arbitrary positions and heights inside the chamber. In order to not degrade the energy of the alpha particles, the chamber was evacuated to a pressure of typically $5\cdot 10^{-2}$~mbar.

\begin{figure}[htbp] 
   \centering
   \includegraphics[width=0.8\textwidth]{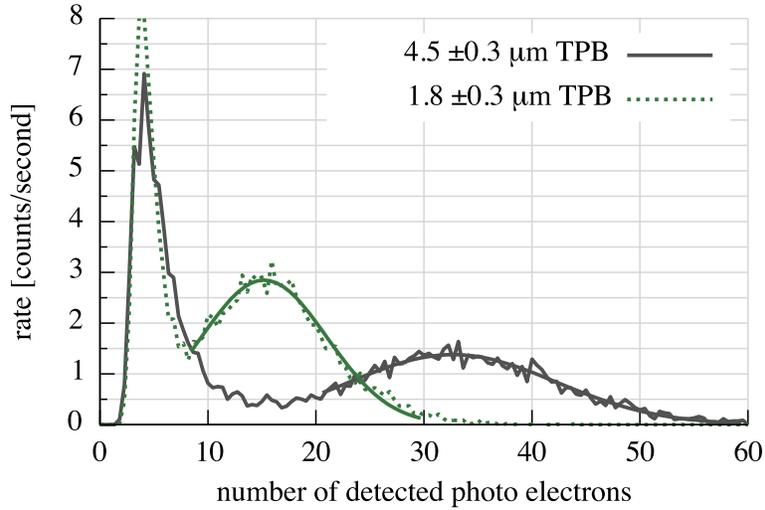} 
   \caption{Background subtracted spectra and gaussian fits for alpha particles going through 4.5$\pm$0.3 $\mu$m (solid dark line )and 1.8$\pm$0.3 $\mu$m (dotted lighter line) of TPB. The peaks at approximately 5 PE are from background.}
   \label{fig:spectrum}
\end{figure}

The light yield was measured with the $^{241}$Am source irradiating different points on the TPB coatings, either pointing straight up or fixed at an angle $\theta$ of 47$^{\circ}$, leading to the alphas travelling distances of 1.8 to 6.5 $\mu$m through the TPB. The PMT was moved for each measurement such that the source was always aligned with the PMT's centre. The PMT signals were read out with a LeCroy WavePro 7100A oscilloscope and the traces were saved to file for offline analysis. Two sample spectra are shown in figure \ref{fig:spectrum}.

The energy loss of the alpha particles going through the TPB was simulated using SRIM\cite{SRIM} tables. The largest and smallest simulated energy loss for the pathlengths studied were 0.18 and 0.69 MeV. The TPB density of 1.16$\pm$0.04~g/cm$^{3}$, found previously, was assumed in the simulation.

\section{Results}
\begin{figure}[htbp] 
   \centering
   \includegraphics[width=0.8\textwidth]{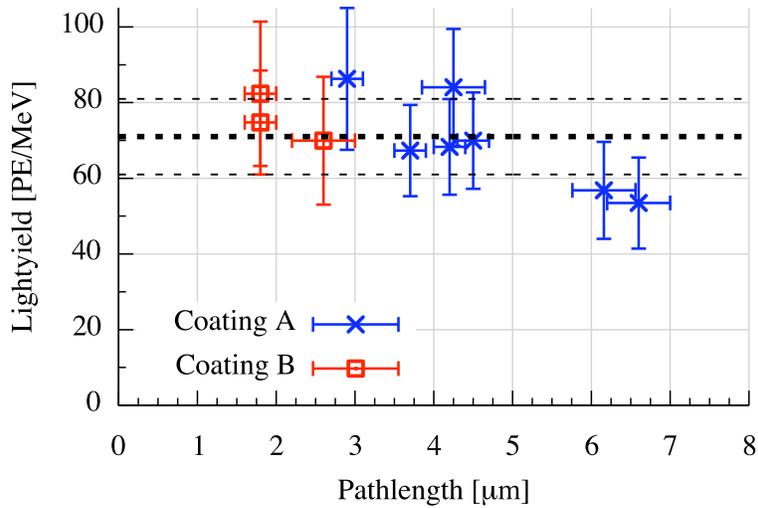} 
   \caption{The measured light yield for different pathlengths the alpha particles had through TPB. The dashed lines represent the mean and standard deviation of the data. The points with the larger error bars on the pathlength variable were measured with the alpha source inclined at 47$^{\circ}$. }
   \label{fig:yield}
\end{figure}

The light yield measured for different pathlengths through the TPB is presented in figure \ref{fig:yield}. The average of these measurements is 71 PE/MeV with a standard deviation of 11 PE/MeV. The geometric acceptance of the PMT is 35\%. Assuming a PMT efficiency of 23\%\cite{datasheet} at 420~nm, the peak wavelength of TPB, 882$\pm$210 photons are emitted per MeV energy deposit. The uncertainty on this number does not include the uncertainty on the PMT efficiency, which is not quoted in the data sheet.

\begin{figure}[htbp] 
   \centering
   \includegraphics[width=0.8\textwidth]{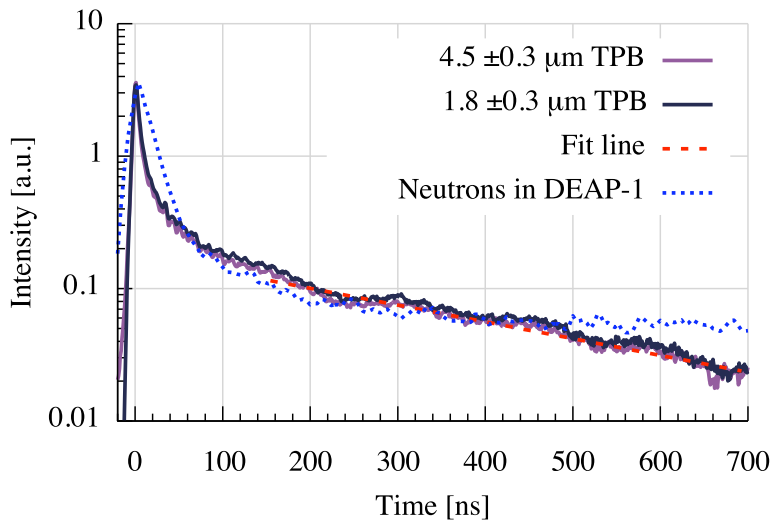} 
   \caption{The solid curves are averaged pulse shapes of alpha induced TPB scintillation for data taken at TPB thicknesses of 1.8~$\mu$m and 4.5~$\mu$m. The dashed line shows best fit to the slow component of those traces. The dotted curve is the average neutron induced liquid argon scintillation pulse shape as measured in DEAP-1. It was normalized to have the same peak height as the TPB pulses.}
   \label{fig:ps}
\end{figure}

The scintillation pulse shape for two measurements is shown in figure \ref{fig:ps}. It consists of a fast and a slow component with lifetimes of 11~$\pm$5~ns and 275~$\pm$10~ns. There are no significant differences in the pulse shape for different TPB thicknesses. 

The so called "fast to total" parameter, the ratio of the pulse intensity in a fast time window and the total pulse intensity, is often used to discriminate electromagnetic interactions in noble liquid detectors from nuclear recoils. Using a fast time window of 50~ns before and 100~ns after the peak position, which is the standard window used in DEAP, and a total integration time of 1~$\mu$s this parameter is 0.67$\pm$0.03 for alpha inducted TPB scintillation.

\section{Discussion}
The theoretical thickness distribution curve describes the measured radial TPB thickness distribution well, even though the pressure in the evaporator as well as the distance between the TPB surface and the substrate change during evaporation. We therefore believe this method of deposition monitor calibration to be reliable for determining the TPB thickness at radii other than where it was directly measured. 

The measured photon yield of 882$\pm$137 photons is a factor of 5 smaller than that measured by Hull et al.\cite{Hull:2009p10088} for alpha scintillation in a TPB monocrystal.
The light yields for coating A and B are consistent within errors, indicating that the pressure in the evaporator during the evaporation process has no significant influence on it. The two data points at 6.1 and 6.6~$\mu$m pathlength seem to be lower than the ones taken at shorter pathlengths. These points were measured with the source inclined by 47$^{\circ}$, so that this effect is not caused by increased absorption in a thicker TPB layer. The other points measured with an inclined source do not show a reduced light yield. It is possible that the energy loss simulation becomes inaccurate for longer pathlengths.
The TPB coatings were frequently exposed to fluorescent lighting in the lab without a noticable change in light yield over 3 months.

The main source of error on the light yield measurement comes from the uncertainty in the energy loss\footnote{The uncertainty in the PMT efficiency is certainly a large source of error as well, but not as relevant for DEAP-1, because the light yield in terms of PE/MeV can be used where different PMTs are compared directly.}. This uncertainty is due to the uncertainty in the relative position between the alpha source and the the axis of the crucible during evaporation, and the uncertainty in the parameter $\frac{u}{\alpha}$ in the radial mass distribution equation. In future work, this error can be significantly reduced by developing a more advanced substrate positioning system in the evaporator and in the scintillation chamber. 

Excitation of TPB with photons excites short lived singlet states\cite{Mckinsey:1997p6265,laser}, which have PMT signals with an Fprompt parameter of 1. The measurement of a 275~ns compontent indicates that alpha particles excite a triplet state, next to a short lived singlet state.  This leads to the lower Fprompt parameter. The measured 11 ns for the singlet state has a large uncertaintly, since the shape is convolved with the time resolution of the PMT. 

The Fprompt parameter of 0.67 measured here is very close to the Fprompt of 0.75 measured in the DEAP-1 detector for neutron events. WIMP and neutron signals are expected to have the same pulse shape and the average neutron pulse shape measured in DEAP-1 is shown in figure \ref{fig:ps} for comparison. TPB scintillation could therefore lead to non-WIMP events in the signal region of that experiment. However, the liquid argon scintillation lifetimes of 6.7~ns for the singlet and 1600~ns\cite{Hitachi:1983p6} for the triplet component are much different from the TPB scintillation lifetimes measured here, suggesting that an improved pulse shape discrimination can be developed. In future work, we plan to study whether the scintillation pulse shape and light yield are temperature dependent.

\section{Conclusion}
We have shown that TPB scintillates under alpha particle excitation with a light yield of 882$\pm$210 photons/MeV and a pulse shape that consists of a fast and a slow component. This measurement will help understand surface alpha backgrounds in experiments using TPB and can be used to evaluate pulse shape discrimination methods for discriminating surface alpha from nuclear recoil events. 

\section{Acknowledgements}
We would like to thank Prof.~Gregory Jerkiewicz and Dr.~Micha\l\ Grde\'n for providing access to the profiler and assistance in using it, as well as Rob Gagnon for technical assistance. This work was supported by the National Science and Engineering Research Council of Canada (NSERC), by the Canada Foundation for Innovation (CFI) and by the Ontario Ministry of Research and Innovation (MRI).

\bibliographystyle{model1-num-names}
\bibliography{library}
\end{document}